\documentclass[12pt,hidelinks,leqno]{article}
\usepackage[english]{babel}
\usepackage{amsmath,amsthm,amssymb}
\usepackage{mathabx}
\usepackage[top=3.5cm, bottom=3.5cm, left=3.3cm, right=3.3cm]{geometry}
\usepackage{graphicx}
\usepackage{amsfonts}
\usepackage[utf8]{inputenc}
\usepackage{tikz}
\usepackage{tikz-cd}
\usetikzlibrary{trees}
\usetikzlibrary{matrix}
\usepackage{verbatim}
\usepackage{appendix}
\usepackage{natbib}
\usepackage{datetime}
\usepackage{setspace}
\usepackage{bookmark}
\usepackage{quoting} 
\usepackage{floatrow}
\usepackage{authblk}

\usepackage{pgfplots}
\usetikzlibrary{patterns}
\pgfplotsset{width=12cm,compat=1.9}
\pgfplotsset{compat = newest}
\usepackage{filecontents}

\usepackage{array}
\usepackage{booktabs}

\usetikzlibrary{decorations.markings}

\newcommand{\bysame}{\leavevmode\hbox to3em{\hrulefill}\,}

\newtheorem{thm}{Theorem}
\newtheorem{cor}{Corollary}
\newtheorem{lem}{Lemma}
\newtheorem{prop}{Proposition}
\newtheorem{claim}{Claim}
\theoremstyle{definition}

\theoremstyle{proof}

\theoremstyle{definition}
\newtheorem{defn}{Definition}
\newtheorem{rem}{Remark}
\numberwithin{equation}{section}

\begin{document}
	\title{Super-Nash Performance\footnote{Department of Political Economy, King's College London, London, UK \\ E-mail: mehmet.ismail@kcl.ac.uk.}}

\author{Mehmet S. Ismail\footnote{
This subsumes my PhD paper, titled ``Maximin equilibrium,'' from Maastricht University. I am grateful to Ronald Peeters and Jean-Jacques Herings for their valuable feedback. I also thank the seminar participants at the Maastricht MLSE Seminar (February 2014), the International Conference on Game Theory (Stony Brook University, 2014), the Paris PhD seminar (IHP), University of S\~{a}o Paulo, Bielefeld University, Saint-Louis University, Brussels (2014), the University of Paris 1 (CES Economic Theory Seminar, 2014), the FUR Conference (Rotterdam University, 2014), the 8th Israeli Game Theory Conference, the 5th World Congress of the Game Theory Society, Aix-Marseille School of Economics (2018), the LSE (STICERD Work in Progress Seminars, 2018), the Lancaster Game Theory Conference, the University of Amsterdam ILLC (COMSOC Seminar, 2019), and SAET 2022, among others, for their valuable comments.}}

	\date{4 March, 2025}
	\maketitle
	\setcounter{footnote}{1}
	\begin{abstract}
		In this paper, I introduce a novel benchmark in games, \textit{super-Nash performance}, and a solution concept, \textit{optimin}, whereby players maximize their minimal payoff under unilateral profitable deviations by other players. Optimin achieves super-Nash performance in that, for every Nash equilibrium, there exists an optimin where each player not only receives but also guarantees super-Nash payoffs under unilateral profitable deviations by others. Further, optimin generalizes Nash equilibrium in $n$-person constant-sum games and coincides with it when $n=2$. Finally, optimin is consistent with the direction of non-Nash deviations in games in which cooperation has been extensively studied.
	\end{abstract}

    

	\doublespacing
	\section{Introduction}
	\label{sec:intro}
	
	Since the early 1990s, artificial intelligence (AI) systems have gradually achieved `superhuman performance' in major competitive games such as checkers, chess, Go, and poker \citep{campbell2002,schaeffer2007,silver2016,brown2019}. These games belong to a class of zero-sum games, where `winning' has a clear definition, and approximating a Nash equilibrium typically yields successful performance against human players. However, most economic and social interactions occur in non-zero-sum settings, where there is a gap in the literature regarding a rigorous, game-theoretical definition of successful performance.

	In this paper, I define a novel benchmark for successful performance in $n$-person games. A strategy profile is said to achieve \textit{super-Nash performance} if it satisfies two conditions: (i) each player receives a super-Nash payoff, i.e., at least as high as a Nash equilibrium payoff, and (ii) each player guarantees a super-Nash payoff under unilateral profitable deviations by other players. However, this guarantee does not extend to all possible deviations. I justify this definition with both theoretical and empirical arguments. First, a `good' measure of performance should not rely solely on the direct payoffs received by the players but should also consider the counterfactual payoffs, i.e., how well these players would perform if they played against players that `exploited' their strategies. For example, two AIs that cooperate in a prisoner's dilemma (PD) would receive a payoff that is strictly greater than their Nash equilibrium payoff. However, they would perform poorly against an opportunistic AI that defects. Second, extensive experimental evidence suggests that humans already achieve super-Nash performance in a variety of non-zero-sum games in which there are gains from cooperation. These include the finitely repeated PD, the finitely repeated public goods game, the centipede game, and the traveler's dilemma \citep{axelrod1980,mckelvey1992,capra1999,goeree2001,rubinstein2007,lugovskyy2017,embrey2017}. In these games, humans not only consistently achieve super-Nash payoffs but also guarantee these super-Nash payoffs even when other players anticipate non-Nash behavior and respond `opportunistically.'
	
	I further introduce a novel solution concept, \textit{optimin}, designed to achieve super-Nash performance in any $n$-person game. This concept extends the maximin strategy concept, introduced by \cite{neumann1928}, from zero-sum to non-zero-sum settings. Informally, an optimin is a strategy profile in which players simultaneously maximize their minimal payoffs under unilateral profitable deviations by other players. Optimin achieves super-Nash performance in every game: for each Nash equilibrium, there exists an optimin where each player not only receives a super-Nash payoff but also guarantees a super-Nash payoff, even if others deviate unilaterally and profitably from the optimin (Proposition~\ref{prop:nashvsoptimin}). However, this guarantee is weaker than that of a maximin strategy, as it holds only under the constraint of profitable deviations. As is well known, a profile of maximin strategies cannot Pareto dominate a Nash equilibrium. Similarly, a Nash equilibrium cannot Pareto dominate an optimin point.
	
	Optimin either coincides with or generalizes well-known concepts in certain sub-domains. For example, optimin coincides with a pair of maximin strategies in potentially infinite two-person zero-sum games, even when a Nash equilibrium does not exist (Proposition~\ref{thm:zerosum}). Furthermore, optimin generalizes Nash equilibrium in $n$-person constant-sum games and coincides with it when $n=2$ (Remark~\ref{rem:constant-sum}). The main theorem states that every $n$-person game possesses an optimin point in possibly mixed strategies (Theorem~\ref{thm:existence}). However, when restricted to pure strategies, a pure strategy optimin point exists in every finite $n$-person game (Remark~\ref{property:ordinal}). In contrast, it is well known that a pure Nash equilibrium does not exist in general and that computing even approximate mixed strategy Nash equilibria is `hard' \citep*{gilboa1989b,daskalakis2009}. While this does not imply that finding a mixed strategy optimin point is computationally more efficient, the pure strategy optimin has an advantage: it allows for focusing solely on the set of pure strategy profiles, which is finite, rather than the infinite set of mixed strategy profiles. This property is particularly useful in large games, where dealing with mixed strategies can be computationally demanding.
	
	While the theoretical properties of optimin appear promising, it is not immediately clear how optimin would perform in experimental games where players consistently exhibit non-Nash behavior yet obtain super-Nash payoffs. In section~\ref{sec:apps}, I analytically derive optimin predictions in four well-studied experimental games---the centipede game, the traveler's dilemma, the finitely repeated $n$-person public goods game, and the finitely repeated PD---in which Nash equilibrium and optimin predictions are in stark contrast. I show that, unlike Nash equilibrium, optimin is consistent with the comparative statics of behavior in these games.
	
	\section{The optimin criterion}
	\label{sec:defn}
	In this section, I begin by introducing the setup for non-cooperative games. Next, I define the concepts of performance and optimin. I then provide an example to illustrate optimin and discuss the intuition behind it. Finally, I explore potential modifications to the optimin criterion.
	
	\subsection{Setup}
	
	Let $\Gamma=(S_i, u_i)_{i\in N}$ be an $n$-person non-cooperative game in mixed extension. Here, $N=\{1,\dots,n\}$ denotes the finite set of players. For each player $i\in N$, $S_i$ is the set of all probability distributions over the finite action set $S'_i$. The set of all mixed strategy profiles is $S=\bigtimes_{i \in N} S_i$. The function $u_i:S\rightarrow \mathbb{R}$ is the von Neumann-Morgenstern expected utility function for each player $i\in N$. A mixed strategy for player $i$ is denoted by $p_i\in S_i$, and a mixed strategy profile by $p\in S$.
	
	I first formally define the performance function and the optimin criterion, and then informally discuss the concept.
	
	\begin{defn}
		\label{def:better_response}
		For each player $i\in N$, the correspondences $B_i(\cdot)$ and $B_{\text{-}i}(\cdot)$ are defined as follows.
		\[
		B_i(p)=\{p'_i\in S_i\mid u_i(p'_i,p_{\text{-}i})>u_i(p_i,p_{\text{-}i})\}\cup \{p_i\}, \text{ and } B_{\text{-}i}(p)=\bigtimes_{j\in N\setminus\{i\}}B_{j}(p).
		\]
	\end{defn}%
	For a given mixed strategy $p$, $B_{\text{-}i}(p)$ represents the set of all profiles that includes both $p$ and all unilateral strictly profitable deviations from $p$ by players other than $i$.\footnote{Note that correlated deviations are not considered in this definition. This possibility is explored in the following subsection.} Next, I define the performance of a strategy profile $p$ for player $i$ as the player's minimal payoff under $B_{\text{-}i}(p)$. While it is not necessary to explicitly define the performance function for the definition of optimin, it is helpful to refer to the performance of any strategy profile, not just optimin, in applications and certain proofs.
	
	\begin{defn}[Performance function]
		\label{def:value}
		For each player $i\in N$, the $i$th component of the (optimin) \textit{performance function} $\pi: S\rightarrow \mathbb{R}^n$ is defined as
		\[
		\pi_i(p)=\inf_{p'_{\text{-}i}\in B_{\text{-}i}(p)} u_{i}(p_i,p'_{\text{-}i}).
		\]
	\end{defn}
	
	The performance of profile $p$ for player $i$ is defined as the minimum payoff the player receives from either (i) $p$ itself, or (ii) any collection of unilateral profitable deviations by other players. As mentioned in the Introduction, this performance depends on not only the direct payoffs but also the counterfactual payoffs a player might receive if other players deviate unilaterally and profitably. The next step involves comparing the performances of different strategy profiles by Pareto optimization of the vector-valued performance function. A strategy profile is called an optimin point in a game if it is Pareto undominated with respect to the performance function.
	
	\begin{defn}
		\label{def:optimin}
		A profile $\bar{p} \in S$ is called an \emph{optimin point}, or said to satisfy the \textit{optimin criterion}, if for every player $i \neq j$ and every $p'\in S$, $\pi_i(p')>\pi_i(\bar{p} )$ implies that there is some player $j$ with $\pi_j(p')<\pi_j(\bar{p})$.
	\end{defn}
	
	Note that, unlike the concept of a maximin strategy, the optimin criterion is a property of a strategy profile, similar to Nash equilibrium. Before discussing the relationship between these concepts, I first illustrate an optimin solution in a game and then provide a more detailed explanation of the intuition behind this definition, along with potential modifications.
	
	\begin{figure}
		\[
		\begin{array}{ r|c|c|c| }
			\multicolumn{1}{r}{}
			&  \multicolumn{1}{c}{\text{Left}}
			& \multicolumn{1}{c}{\text{Center}}
			& \multicolumn{1}{c}{\text{Right}} \\
			\cline{2-4}
			\text{Top} &  100,100 & 100,105 & 0,0 \\
			\cline{2-4}
			\text{Middle}&  105,100 & 95,95 & 0,210  \\
			\cline{2-4}
			\text{Bottom}&  0,0 & 210,0 & 5,5\\
			\cline{2-4}
		\end{array}
		\qquad
		\begin{array}{ r|c|c|c| }
			\multicolumn{1}{r}{}
			&  \multicolumn{1}{c}{\text{Left}}
			& \multicolumn{1}{c}{\text{Center}}
			& \multicolumn{1}{c}{\text{Right}} \\
			\cline{2-4}
			\text{Top} &  (100,100) & (100,0) & (0,0) \\
			\cline{2-4}
			\text{Middle}&  (0,100) & (0,0) & (0,5)  \\
			\cline{2-4}
			\text{Bottom}&  (0,0) & (5,0) & (5,5)\\
			\cline{2-4}
		\end{array}
		\]
		\caption{An illustrative game (left) and its performance function (right).}
		\floatfoot{Note: The unique optimin point is (Top, Left), whereas every strategy is a maximin strategy in this game. The unique Nash equilibrium is (Bottom, Right).}
		\label{fig:illustrative}
	\end{figure}
	
	\subsubsection*{Illustrative example} 
	Consider the game in Figure~\ref{fig:illustrative} (left), where, for simplicity, attention is restricted to pure strategies. Notice that every action is a maximin strategy, guaranteeing a payoff of 0. Now, consider the strategy profile (Top, Left). Even if player 2 (he) has a profitable deviation to `Center', player 1 (she), who plays Top, would still secure 100 at profile (Top, Center). Although player 2 could also deviate to `Right'---a viable choice under the maximin strategy concept---such a deviation is implausible, as he would incur a substantial loss. Thus, player 1's performance, or minimal payoff under the unilateral profitable deviation of the opponent, at the profile  (Top, Left) is 100. Since the game is symmetric, player 2's performance at this profile is also 100. As illustrated in Figure~\ref{fig:illustrative} (right), (Top, Left) uniquely maximizes the performance function, and hence it is the optimin point. 
	
However, note that although the performance function has a global maximum (i.e., the unique Pareto undominated profile) in this game, it does not always have one in general. Hence, there may be more than one optimin in some games. A simple example is the `Battle of the Sexes,' where there are two optimin points, (Football, Football) and (Opera, Opera), and each player strictly prefers one over the other. A typical payoff matrix for this game is presented below. 
	\[
	\begin{array}{ r|c|c| }
		\multicolumn{1}{r}{}
		&  \multicolumn{1}{c}{\text{Football}}
		& \multicolumn{1}{c}{ \text{$\;$ Opera $\;$} }\\
		\cline{2-3}
		\text{Football}&  3,2 & 0,0 \\
		\cline{2-3}
		\text{Opera}&  0,0 & 2,3 \\
		\cline{2-3}
	\end{array}
	\]
	
	Returning to the $3\times 3$ game, consider the strategy profile (Middle, Center), which might initially seem appealing. However, the performance of this profile for player 1 is 0, as player 2 could unilaterally and profitably deviate to `Right,' leaving player 1 with a payoff of 0. For similar reasons, player 2's performance at this profile is also 0. As a final example, each player's performance at the Nash equilibrium (Bottom, Right) is 5, since no player has a unilateral profitable deviation from it.
	
	\subsubsection*{Informal discussion}
	
	In plain words, a strategy profile $\bar{p}$ is called an optimin point if players simultaneously maximize their minimal payoffs, subject to the constraint of unilateral profitable deviations (i.e., opportunistic behavior) by other players. If $\bar{p}$ is a Nash equilibrium, no unilateral profitable deviation is possible. However, if $\bar{p}$ is not a Nash equilibrium, at least one player (say, $j$) has a unilateral profitable deviation from $\bar{p}$. Despite this, the other players, who follow their part of the profile $\bar{p}$, are guaranteed specific payoffs, even if player $j$ deviates unilaterally and profitably from $\bar{p}$. That being said, they are not guaranteed any specific payoffs if player $j$ deviates non-profitably from $\bar{p}$.
	
	It is also informative to differentiate between the optimin and maximin concepts. Under a maximin strategy, a player chooses a strategy to maximize the minimum utility, considering all potential choices by other players, even those that may result in a significantly negative outcome for the deviating players. In contrast, under an optimin point, $\bar{p}$, a player is restricted to making only profitable deviations from $\bar{p}$.
	
	Note that the optimin criterion assumes fully non-cooperative play: ``each player acts independently without collaboration with any of the others'' \citep{nash1951}. A player is allowed to make any unilateral profitable deviation, without considering how it affects the payoffs of other players. Coalitional or correlated profitable deviations are not considered in the benchmark definition of optimin, although the definition can be modified to include such deviations. I explore a definition of optimin in cooperative games in a companion paper.
	
	\subsection{Potential modifications}
	
	In the definition of optimin, one might also consider types of deviations other than strict better-response deviations, such as weak better-response and best-response deviations. Here, I present an alternative definition of the optimin concept to account for these types of deviations, which may yield implications different from those discussed in this paper.

	Let $\hat{B}_{\text{-}i}(p)$ denote the set of beliefs held by player $i$ regarding other players' potential deviations from $p$, which may be correlated.\footnote{See \citet{ritzberger1995} for a  definition of `behavior correspondences,' which generalizes best-response and better-response correspondences.} Let $\hat{\pi}$ be the $i$th component of the \textit{general performance function} defined as follows: for all $p\in S$ and all $i\in N$, $ \hat{\pi}_i(p)=\inf_{p'_{\text{-}i}\in \hat{B}_{\text{-}i}(p)} u_{i}(p_i,p'_{\text{-}i})$. The general performance of profile $p$ for player $i$ is basically the performance of $i$ at $p$, but the infimum is taken over a different domain. Accordingly, a strategy profile may be called an optimin point if it maximizes this general performance function.
	
	Note that the performance function in Definition~\ref{def:value} assigns each player a unique performance for every strategy profile $p$. This definition can be interpreted as follows. Suppose that, before playing a game, players can make a tacit (non-binding) agreement to play a strategy profile $p$. Then, each player chooses a strategy simultaneously and independently, retaining the option to either follow or deviate from this non-binding agreement. The optimin performance assumes that players evaluate such a profile cautiously, ruling out implausible deviations---those that do not strictly improve the payoff of the deviator while holding the strategies of others fixed. Accordingly, a player's performance from following this agreement is defined as the minimum utility the player receives either (i) from the agreement itself or (ii) under the constraint defined by $\hat{B}_{\text{-}i}$, which is the set of (possibly correlated) beliefs held by player $i$ about potential deviations by other players. In the benchmark definition,  $\hat{B}_{\text{-}i}=B_{\text{-}i}$; however, $\hat{B}_{\text{-}i}$ may also be defined as the weak better-response correspondence or the best-response correspondence. Regardless of the specification, Definitions 1--3 remain well-defined under different choices for $\hat{B}_{\text{-}i}$.
	
	A major implication of defining $B_{\text{-}i}$ using weak better-responses rather than strict better-responses would be that the performance of a Nash equilibrium would not necessarily equal its payoff. While there is no unilateral profitable deviation from a Nash equilibrium, there may be a unilateral weakly profitable deviation that gives the same payoff to the deviator as the Nash equilibrium. Under such weakly profitable deviations, the non-deviator's payoff may be strictly lower than the Nash equilibrium payoff. Thus, if weakly profitable deviations are included in the definition of the performance function, the performance of a Nash equilibrium would generally not equal its payoff. Nevertheless, as in the benchmark definition of $B_{\text{-}i}$, it is arguably more natural for players' performance at a Nash equilibrium to equal their Nash equilibrium payoff.
	
	The optimin criterion essentially applies a two-step evaluation and comparison method for assessing the performance of strategy profiles in games. As described above, one may interpret $\hat{B}_j(p)$ as being the belief of some player $i$ about player $j$'s (or a coalition's) potential deviations, and evaluate the performance of strategy profiles based on that definition. While the benchmark performance function evaluates strategy profiles based on their minimal payoffs, it is conceivable to use alternative evaluation criteria, such as minimax regret \citep{savage1954} or the Hurwicz criterion \citep{hurwicz1951}. Nevertheless, the current definition, which focuses on minimal payoffs, is sufficient to account for some well-documented non-Nash deviations observed in experimental games, as discussed in section~\ref{sec:apps}. Following the evaluation step, the next step involves comparing the performances of different strategy profiles. This is achieved by maximizing the multi-variable performance function via multi-objective maximization or Pareto optimization.\footnote{In \citet{ismail2014}, I called $\bar{p}$ a `maximin equilibrium' if either (i) it is an optimin point, or (ii) for every $i$, $\bar{p}_i \in \arg \max_{q_i\in S_i} \inf_{p'_{\text{-}i}\in B_{\text{-}i}(q_i,\bar{p}_{\text{-}i})} u_{i}(q_i,p'_{\text{-}i})$, which could be thought of as the `equilibrium' counterpart of the optimin.} However, there is no logical relationship between the optimin criterion and Pareto optimality in games. For example, as mentioned earlier, in the Battle of the Sexes game, there are two optimin points, both of which are Pareto optimal. However, an optimin point may be Pareto dominated, as in the PD, where the unique optimin point is (Defect, Defect).
	
	\section{Results}
	
	In this section, I present the main existence result for $n$-person games as well as several useful properties of optimin in different sub-domains.
	
	\subsection{The existence of optimin}
	
	The following theorem establishes the existence of optimin points in mixed strategies.	Its proof relies on Lemma~\ref{value_usc}, which in turn relies on a theorem by  \citet{berge1959}, both of which are provided in the Appendix.
	
	\begin{thm} [Existence]
		\label{thm:existence}
		Let $\Gamma=(S_i, u_i)_{i\in N}$ be a finite $n$-person game in mixed extension. Then, $\Gamma$ possesses an optimin point in possibly mixed strategies.
	\end{thm}
	
	\begin{proof} 
		The proof strategy involves iteratively maximizing each player's performance function over a compact subset of strategy profiles. The upper semi-continuity of performance functions ensures that the sets of maximizers are compact and nonempty. Each subsequent player's performance function is then maximized over the compact domain determined by the previous player's set of maximizers. This iterative process leads to a nonempty compact set of maximizers, which are optimin points.
		
		For $i\in \{1,2\}$, let $\pi_i$ be player $i$'s performance function in a two-player game $(S_i, u_i)_{i\in \{1,2\}}$. Define
		\[
		\pi^{max}_i=\arg\max_{q\in S}\pi_i(q),
		\]
		which is a nonempty compact set because $S$ is compact, and $\pi_i$ is upper semi-continuous by Lemma~\ref{value_usc}. Since $\pi^{max}_i$ is compact and $\pi_j$, where $j\neq i$, is also upper semi-continuous by Lemma~\ref{value_usc}, the set,
		\[
		\pi^{max}_{ij}=\arg\max_{q\in \pi^{max}_i}\pi_j(q),
		\]
		must be nonempty and compact. By construction of the set $\pi^{max}_{ij}$, a strategy profile $\bar{p}\in \pi^{max}_{ij}$ cannot be Pareto dominated by another strategy profile $p\in S$ with respect to the performance function. This implies that $\pi^{max}_{ij}$ is a nonempty compact subset of the set of optimin points in the game. Similarly, the set $\pi^{max}_{ji}$ is also a nonempty compact subset of the set of optimin points. These arguments extend straightforwardly to games with any finite number of players.
	\end{proof}
	
	Neither the convexity of the strategy sets nor the concavity of the utility functions is used in the proof of Lemma~\ref{value_usc} or Theorem \ref{thm:existence}. Thus, the result can be generalized: any game with continuous utility functions and compact strategy spaces possesses an optimin point.
	
	\subsection{Properties of optimin}
	\label{subsec:properties}
	
	First, I show that for every Nash equilibrium, there is an optimin point where both the performance and the actual payoff of the players are weakly greater than their Nash equilibrium payoff.
	
	\begin{prop}[Super-Nash performance]
		\label{prop:nashvsoptimin}
		For every Nash equilibrium $q^*\in S$ in a game $\Gamma$, there exists an optimin point $\bar{p}\in S$ such that for every player $i$,
		\[
		\pi_{i}(\bar{p})\geq u_i(q^*).
		\]
		Moreover, there is no game in which a Nash equilibrium Pareto dominates an optimin point.
	\end{prop}
	
	\begin{proof}
		The proof proceeds by contradiction: assume the negation of the first statement and show that it leads to a contradiction, thereby proving the statement true.
		
		Suppose the negation holds: there exists a Nash equilibrium $q^*\in S$ such that, for every optimin $\hat{p}\in S$, there exists a player $j$, which may depend on $\hat{p}$, such that 
		\begin{equation}
			\label{eq:super-nash}
			\pi_{j}(\hat{p})< u_j(q^*).
		\end{equation}
		In other words, this negation states that for every optimin $\hat{p}$, one can always find a player $j$ such that $j$'s payoff at the Nash equilibrium is strictly greater than $j$'s performance at the optimin point. 
		
		There are two cases to consider: (i) $q^*$ is an optimin point, and (ii) $q^*$ is not an optimin point. Case (i): assume that $q^*$ is an optimin point. Then, for $\hat{p}=q^*$ Inequality~\ref{eq:super-nash} becomes
		\begin{equation}
			\label{eq:super-nash1}
			\pi_{j}(q^*)< u_j(q^*).
		\end{equation}
		
		Since $q^*$ is a Nash equilibrium, for every player $i$,
		\begin{equation}
			\label{eq:super-nash_performance}
			\pi_{i}(q^*) = u_i(q^*),
		\end{equation}
		which means that each player's performance at the Nash equilibrium equals the player's Nash equilibrium payoff, as there is no unilateral profitable deviation. Substituting $j$ for $i$ in Equation~\ref{eq:super-nash_performance}, Inequality~\ref{eq:super-nash1} reduces to $u_j(q^*) < u_j(q^*)$, which is a contradiction.
		
		Case (ii): assume that $q^*$ is not an optimin point. Using the fact that $u_j(q^*) = \pi_{j}(q^*)$, Inequality~\ref{eq:super-nash} implies that
		\begin{equation}
			\label{eq:super-nash2}
			\pi_{j}(\hat{p})<  \pi_{j}(q^*).
		\end{equation}
		This implies that for every optimin $\hat{p}$ (by Theorem~\ref{thm:existence}, at least one exists), one can always find a player $j$ whose performance at the Nash equilibrium $q^*$ is strictly greater than $j$'s performance at the optimin point. However, this would mean that the performance of $q^*$ cannot be Pareto dominated. By definition, $q^*$ must be an optimin point, which contradicts the assumption that $q^*$ is not an optimin point. Therefore, the negation of the first statement in the proposition leads to a contradiction, and hence the statement holds.
		
		Next, I prove the second statement in the proposition: there is no game in which a Nash equilibrium Pareto dominates an optimin point. To reach a contradiction, suppose that there exists a game with a Nash equilibrium $q^*$ that Pareto dominates the optimin point $\hat{p}$: for every player $i$, $u_i(q^*) \geq u_i(\hat{p})$, with at least one strict inequality. Since $q^*$ is a Nash equilibrium, by Equation~\ref{eq:super-nash_performance}, for every player $i$, $\pi_{i}(q^*) = u_i(q^*)$. This equality implies that for every player $i$,
		\begin{equation}
			\label{eq:super-nash3}
			\pi_{i}(q^*) \geq u_i(\hat{p}),
		\end{equation}
		with at least one strict inequality. 
		
		Note that for every player $i$ and every strategy profile $p\in S$, $u_i(p)  \geq \pi_{i}(p)$. In other words, a player's performance at a strategy profile is always weakly less than the player's payoff. Then, Inequality~\ref{eq:super-nash3} implies that
		\[
		\pi_{i}(q^*) \geq \pi_{i}(\hat{p}),
		\]
		with at least one strict inequality. However, this would mean that the performance of $\hat{p}$ is Pareto dominated, contradicting the supposition that $\hat{p}$ is an optimin point. Thus, the second statement in the proposition holds as well.
	\end{proof}
	
	In other words, if the Nash equilibrium $q^*$ is an optimin point, then the inequality in the Proposition~\ref{prop:nashvsoptimin} holds in part because, at a Nash equilibrium, players' performance is equal to their Nash equilibrium payoff. This also explains why a Nash equilibrium cannot Pareto dominate an optimin point: if it did, the performance of the optimin point would be Pareto dominated, leading to a contradiction. Conversely, if the profile $q^*$ is not an optimin point, then by Theorem~\ref{thm:existence} and the definition of optimin, there must exist an optimin such that players' performance is at least as high as their performance at the Nash equilibrium, with at least one player having a strictly higher performance. 
	
	Note that since players' performance is always weakly less than their actual payoff, Proposition~\ref{prop:nashvsoptimin} implies that for every Nash equilibrium, there is an optimin point where players receive a weakly greater payoff than their Nash equilibrium payoff. In addition, Proposition~\ref{prop:nashvsoptimin} would remain valid if the Nash equilibrium is replaced by a profile of maximin strategies in an $n$-person game. This is due to the well-known fact that a player's Nash equilibrium payoff cannot be strictly lower than the player's maximin strategy payoff. 
	
	Recall that the game $\Gamma=(S_i, u_i)_{i\in N}$ is an $n$-person game in mixed extension and that $S'_i$ is the set of all pure strategies for player $i$. Define $G=(S'_i, u'_i)_{i\in N}$ as the game $\Gamma$ restricted to pure strategies. In other words, $\Gamma$ is a mixed extension of $G$.  The following remark shows the existence of an optimin point in pure strategies in $G$.
	
	\begin{rem}
		\label{property:ordinal}
		Every finite $n$-person game $G$ with pure strategies has an optimin point in pure strategies.
	\end{rem}
	
	The proof is straightforward: since there are finitely many pure strategies in finite games, the performance function always has a nonempty set of Pareto optimal points. The existence of optimin points in games restricted to pure strategies is particularly useful in economic applications where finding mixed strategy equilibria can be tedious.

The following remark echoes Proposition~\ref{prop:nashvsoptimin} in the context of games restricted to pure strategies.
	
		\begin{rem}
		\label{property:pure_super-Nash}
		Consider a finite $n$-person game $G$ with pure strategies. If a pure Nash equilibrium exists, then for every such equilibrium, there is a pure optimin point where both the payoffs and the performances of the players are weakly greater than their pure Nash equilibrium payoffs.
	\end{rem}
	
	In words, optimin achieves super-Nash performance in games restricted to pure strategies. The proof follows directly from the proof of Proposition~\ref{prop:nashvsoptimin}, with mixed strategies replaced by pure strategies.
	
	However, an optimin in $G$ does not necessarily achieve super-Nash performance in the mixed extension $\Gamma$. This occurs for two reasons. First, allowing mixed strategies introduces the possibility of mixed Nash equilibria, which may Pareto dominate the performance of an optimin in $G$. Second, an optimin in $G$ is not guaranteed to remain an optimin in $\Gamma$, just as a maximin strategy in $G$ is not necessarily a maximin strategy in $\Gamma$. 

	The following remark provides a simple modification of optimin which coincides with the maximin strategies in $n$-person games. 
	
	\begin{rem}
		\label{prop:reduction}
		Assume that in the definition of the performance function, $B_{\text{-}i}$ is replaced with $S_{\text{-}i}$. Under this modification, an optimin reduces to a profile of maximin strategies.
	\end{rem}
	
	\begin{proof} 
		Taking the infimum over $S_{\text{-}i}$ instead of $B_{\text{-}i}$ in the definition of optimin implies that for every player $i$, 
		\[	
		\max_{p\in S} \inf_{p'_{\text{-}i}\in B_{\text{-}i}(p)} u_{i}(p_i,p'_{\text{-}i}) = \max_{p\in S} \min_{p'_{\text{-}i}\in S_{\text{-}i}} u_{i}(p_i,p'_{\text{-}i}).
		\]
		Thus, if $\bar{p}$ is an optimin point under this definition, then for every player $i$, $\bar{p}_i$ is a maximin strategy.
	\end{proof}
	
	\citet[p. 70]{harsanyi1988} argue that invariance with respect to positive affine transformations of the utilities is a fundamental requirement for a solution concept. This requirement is satisfied by the optimin criterion, as the following proposition shows.
	
\begin{rem}
\label{invariance}
Optimin points are invariant under positive affine transformations of utilities, which may differ for each player.
\end{rem}

\begin{proof}
Let $\Gamma=(S_i, u_i)_{i\in N}$ and $\hat{\Gamma}=(S_i, \hat{u}_i)_{i\in N}$ be two games that differ only as follows. For some player $j$, $\hat{u}_{j}(\cdot)= \alpha_j u_{j}(\cdot)+ \beta_j$, where $\alpha_j>0$ and $\beta_j$ is a constant. For every player $i\neq j$, $u_i(\cdot)=\hat{u}_i(\cdot)$. In words, player $j$'s utility function $\hat{u}_j$ in $\hat{\Gamma}$ is a positive affine transformation of $u_j$.

The sets of optimin points for $\Gamma$ and $\hat{\Gamma}$ are the same because:  
(i) the strict better-response correspondence remains unchanged under positive affine transformations;  
(ii) $\alpha$ and $\beta$ can be factored out of the infimum in Definition~\ref{def:value} without affecting the set of minimizers; and  
(iii) Pareto optimal points are invariant under positive affine transformations.

These arguments (i)--(iii) extend to positive affine transformations of any player's utilities, even when different positive affine transformations are applied to the utilities of different players.
\end{proof}

	The following remark illustrates that the optimin criterion in constant-sum games generalizes Nash equilibrium.
	
	\begin{rem}
		\label{rem:constant-sum}
		Every Nash equilibrium in an $n$-person constant-sum game $\Gamma$ is an optimin point.
	\end{rem}
	
	\begin{proof}
		Let $q^*$ be a Nash equilibrium in an $n$-person constant-sum game. As noted earlier, for every player $i$, $\pi_{i}(q^*) = u_i(q^*)$, meaning that player $i$'s performance is equal to the Nash equilibrium payoff. Since every strategy profile is Pareto optimal in $n$-person constant-sum games, the performance of the Nash equilibrium $q^*$ is also Pareto optimal. Thus, $q^*$ is an optimin point.
	\end{proof}
	
	As shown in the following subsection, in finite two-person zero-sum games, a strategy profile is a Nash equilibrium if and only if it is an optimin point.
	
	\subsection{Zero-sum games}
	\label{sec:wald}
	
	A two-person zero-sum game is denoted by the tuple $(Y_1,Y_2,u_{1},u_{2})$ where $Y_1$ and $Y_2$ denote the (not necessarily finite) sets of pure strategies for the two players. For every strategy profile $y\in Y = Y_1\times Y_2$, the utility functions satisfy $u_1(y)+u_2(y)=0$. The definitions of the performance function and optimin are extended to zero-sum games with possibly infinite strategy sets as follows: for every player $i$, replace the set of strategies $S_i$ with $Y_i$ and the set of strategy profiles $S$ with $Y$ in Definitions \ref{def:better_response} and \ref{def:optimin}.
	
	In zero-sum games, the following proposition shows that a strategy profile is an optimin point if and only if it is a pair of maximin strategies \citep{neumann1928}.
	
	\begin{prop}
		\label{thm:zerosum}
		Let $(Y_1,Y_2,u_{1},u_{2})$ be a two-person zero-sum game. A strategy profile $(y^*_1,y^*_2)\in Y$ is an optimin point if and only if for every player $i\in \{1,2\}$,
		\[
		y^*_i\in\arg\max_{y_i\in Y_i}\inf_{y_{\text{-}i}\in Y_{\text{-}i}}u_i(y_i,y_{\text{-}i}).
		\]
	\end{prop}
	
	\begin{proof} I begin by showing that if a strategy profile $y^*$ is an optimin point, then each player's strategy $y^*_i$, where $i\in \{1,2\}$, is a maximin strategy in the game $(Y_1,Y_2,u_{1},u_{2})$. 
		
		The proof proceeds in two steps. First, I show that the performance function simplifies to the desired infimum. Second, I prove that the performance of an optimin point must be Pareto dominant in a zero-sum game. This ensures that each strategy in the optimin profile separately maximizes the minimum utility over the opponent's entire strategy set.
		
		Step 1: I show that for every player $j \neq i$, $B_j$ can be replaced with $Y_j$ in the definition of player $i$'s performance function. In other words, for every $(y_i,y_j)\in Y$, the performance function simplifies as follows: 
		\[
		\pi_i(y_i,y_j)=\inf_{y'_{j}\in B_{j}(y)} u_{i}(y_i,y'_{j}) = \inf_{y'_j\in Y_j} u_{i}(y_i,y'_{j}).
		\]
		
		There are two cases to consider. Case 1: assume that there exists a strategy $\bar{y}_j\in Y_j$ such that $\bar{y}_j\in\arg\min_{y'_j\in Y_j}u_i(y_i,y'_j)$, i.e., $\bar{y}_j$ minimizes $i$'s payoff. This implies that 
		\[
		\pi_i(y_i,y_j)=\min_{y'_j\in Y_j}u_i(y_i,y'_j)=u_i(y_i,\bar{y}_j),
		\]
		as desired. 
		
		Case 2: assume that for all $y'_j\in Y_j$, there exists $y''_j\in Y_j$ such that $u_i(y_i,y''_j)<u_i(y_i,y'_j)$. This implies that 
		\[
		\pi_i(y_i,y_j)=\inf_{y'_j:u_{i}(y_i,y'_{j})<u_{i}(y_i,y_{j})} u_{i}(y_i,y'_{j})=\inf_{y'_j\in Y_j} u_{i}(y_i,y'_{j}).
		\]
		
		Step 2: I next show that the performance of the optimin point $(y^*_{i},y^*_{j})$ must be Pareto dominant in the game. By contraposition, suppose that it is not Pareto dominant; that is, there is another optimin point $(\hat{y}_{i},\hat{y}_{j})$ such that for $i\neq j$, 
		\[
		\pi_i(y^*_{i},y^*_{j})>\pi_i(\hat{y}_{i},\hat{y}_{j}) ~\text{ and }~ \pi_j(y^*_{i},y^*_{j})<\pi_j(\hat{y}_{i},\hat{y}_{j}). 
		\]
		Then, by Step 1, $\pi_{i}(y^*_{i},y^*_{j})=\pi_{i}(y^*_{i},\hat{y}_{j})$ and $\pi_{j}(\hat{y}_{i},\hat{y}_{j})=\pi_{j}(y^*_{i},\hat{y}_{j})$. This implies that the performance of $(y^*_i,\hat{y}_j)$ Pareto dominates the performance of $(y^*_{i},y^*_{j})$, which contradicts the supposition that $(y^*_{i},y^*_{j})$ is an optimin point. Therefore, the performance of $(y^*_{i},y^*_{j})$ is Pareto dominant with respect to the performance function. This shows that each strategy of the optimin point $y^*$ separately maximizes the minimum utility over the opponent's strategy set.
		
		Conversely, I show that if each player's strategy is a maximin strategy, then the pair of maximin strategies is an optimin point. Assume that for each $i\neq j$, $y^*_i$ is a maximin strategy:
		\[
		y^*_i\in\arg\max_{y_i\in Y_i}\inf_{y_j\in Y_j}u_i(y_i,y_j).
		\]
		This implies that for all $(y'_{i},y'_{j})\in Y$ and for every $i$, $\pi_i(y^*_{i},y^*_{j})\geq \pi_i(y'_{i},y'_{j})$. Hence, the performance of $(y^*_{i},y^*_{j})$ is Pareto dominant, which implies that it is an optimin point.
	\end{proof}
	
	\begin{cor}
		\label{cor:zerosum}
		Let $\Gamma=(S_i, u_i)_{i\in \{1,2\}}$ be a finite two-person zero-sum game in mixed extension. The set of optimin points in $\Gamma$ coincides with the set of Nash equilibria. 
	\end{cor}
	
	Proposition~\ref{thm:zerosum} shows that an optimin point always corresponds to a pair of maximin strategies, even in games that are not necessarily finite, where a Nash equilibrium may not exist. However, in finite zero-sum games, this result implies that the set of optimin points coincides with the set of Nash equilibria, which is always nonempty.
	
	The zero-sum setup discussed in this subsection is also applicable to Wald's (\citeyear{wald1950}) decision theory, which is based on the idea that a statistician should use a maximin strategy to minimize the maximum risk in a carefully constructed game against Nature. In this decision-making problem under uncertainty, Nature is assumed to maximize the risk, which makes the game between the statistician and Nature a zero-sum game.

	\section{Applications: optimin and non-Nash behavior}
	\label{sec:apps}
	
	Proposition~\ref{prop:nashvsoptimin} shows that for every Nash equilibrium, there is an optimin point where every player guarantees a super-Nash payoff under the opportunistic behavior of others. Although this theoretical guarantee is weaker than that of a maximin strategy, it seems promising. However, it is not immediately clear how different the predictions of optimin and Nash equilibrium can be. 
	
	In this section, I apply optimin to four well-studied games in experimental economics: the centipede game, the traveler's dilemma, the finitely repeated PD, and the finitely repeated $n$-person public goods game, where the predictions of Nash equilibrium and optimin are in sharp contrast.
	
	\subsection{The centipede game}
	\label{sec:centipede}
	
	Let the tuple $(\{1, 2\},$ $\{C,S\}, u_1, u_2, m)$ represent a centipede game, which is a two-person extensive-form game of perfect information where each player $i$ can choose either $C$ (continue) or $S$ (stop) at each decision node \citep{rosenthal1981}. The game consists of $m$ decision nodes, where $m$ is finite and even, and the players alternate actions, with player 1 choosing at the first node. Player 1 is active at odd decision nodes, $1,3,\dots,m-1$, and player 2 is active at even nodes, $2,4,\dots,m$.
	Aumann's (\citeyear{aumann1998}) version of this game is defined by the following payoff function, illustrated in Figure~\ref{fig:payoffs_centipede}.

	\[
	u_1(s_i, s_j),u_2(s_i, s_j)\;=\;
	\begin{cases}
		k+1,\;k
		& \text{if the first $S$ is chosen at an odd node }k,\\
		k-1,\;k+2
		& \text{if the first $S$ is chosen at an even node }k,\\
		m+2,\;m+1
		& \text{if } \text{no $S$ is chosen.}
	\end{cases}
	\]

	\begin{figure}
		\centering
		\begin{tikzpicture}[font=\footnotesize,scale=1.05]
			\tikzstyle{solid node}=[circle,draw,inner sep=1.2,fill=black];
			\tikzstyle{hollow node}=[circle,draw,inner sep=1.2,fill=black];
			
			\node(0)[solid node]{}
			child[grow=down]{node{} edge from parent node[left]{$S$}}
			child[grow=right]{node(1)[solid node]{}
				child[grow=down]{node{} edge from parent node[left]{$S$}}
				child[grow=right]{node(2)[solid node]{}
					child[grow=down]{node{} edge from parent node[left]{$S$}}
					child[grow=right]{node(3)[solid node]{}
						child[grow=down]{node{} edge from parent node[left]{$S$}}
						child[grow=right]{node(4)[solid node]{}
							child[grow=down]{node{} edge from parent node[left]{$S$}}
							child[grow=right]{node(5){} 
								edge from parent node[above]{$C$}
							}
							edge from parent node[above]{$\cdots$}
						}
						edge from parent node[above]{$C$}
					}
					edge from parent node[above]{$\cdots$}
				}
				edge from parent node[above]{$C$}
			};
			
			\foreach \x in {0,2} \node[above] at (\x){1};
			\foreach \x in {1,3} \node[above] at (\x){2};
			\node[above] at (4){2};
			
			\node[below] at (0-1){$\begin{pmatrix}2 \\ 1\end{pmatrix}$};
			\node[below] at (1-1){$\begin{pmatrix}1 \\ 4\end{pmatrix}$};
			\node[below] at (2-1){$\begin{pmatrix}k_{\text{o}}+1 \\ k_{\text{o}}\end{pmatrix}$};
			\node[below] at (3-1){$\begin{pmatrix}k_{\text{e}}-1 \\ k_{\text{e}}+2\end{pmatrix}$};
			\node[right] at (5){$\begin{pmatrix}m+2 \\ m+1\end{pmatrix}$}; 
			\node[below] at (4-1){$\begin{pmatrix}m-1 \\ m+2\end{pmatrix}$}; 
			
		\end{tikzpicture}
		\caption{Centipede game payoff function. The values $k_{\text{o}}$ and $k_{\text{e}}$ represent the nodes when $k$ is odd and even, respectively.}
		\label{fig:payoffs_centipede}
	\end{figure}
	
	\begin{claim}
		\label{prop:centipede}
		The performance function of the centipede game $(\{1, 2\},$ $\{C,S\}, u_1, u_2, m)$ is given as follows:
		\[
		\pi_1(s_1, s_2), \pi_2(s_1, s_2) \;=\;
		\]
		\[
		\begin{cases}
			2,\;1
			& \text{if the first $S$ is chosen at node }k=1,\\
			k-2,\;k
			& \text{if the first $S$ is chosen at an odd node }k>1,\\
			k-1,\;k-1
			& \text{if the first $S$ is chosen at an even node }k,\\
			m-1,\;m+1
			&  \text{if } \text{no $S$ is chosen.}
		\end{cases}
		\]
	\end{claim}
	
	\begin{proof}
		First, consider a profile in which player 1 chooses $S$ at the first node. Since neither player has a (unilateral) profitable deviation from this profile that decreases the other's payoff, the players' performances are equal to their individual payoffs. 
		
		Second, consider a strategy profile where player 1 chooses $S$ first (i.e., prior to player 2) at an odd node $k>1$. Player 1 has no profitable deviation that decreases player 2's payoff. Thus, player 2's performance is equal to player 2's own payoff of $k$. By contrast, as long as $k>1$, player 2 has a unique profitable deviation to choose $S$ at node $k-1$, which undercuts player 1, in which case player 1's worst-case payoff would decrease to $k-2$. 
		
		Third, consider a strategy profile where player 2 chooses $S$ first at an even node $k$. Similar to the previous case, player 2 has no profitable deviation that decreases player 1's payoff. Thus, player 1's performance is equal to player 1's own payoff $k-1$ at this profile. However, player 1 has a profitable deviation to choose $S$ at node $k-1$, undercutting player 2. As a result of this deviation, player 2's payoff would decrease to $k-1$. 
		
		Finally, consider the case of the strategy profile where both players always choose $C$. Player 1 has no profitable deviation, while player 2 has a profitable deviation to $S$ at node $m$, which decreases player 1's payoff to $m-1$. Therefore, the players' performances at this node are $(m-1, m+1)$.
	\end{proof}
	
	Given the performance function, it is clear that the cooperative strategy profile of always choosing $C$ is the unique optimin when $k\geq 4$ because its performance, $(m-1, m+1)$, Pareto dominates the performance of all other profiles.\footnote{Aumann's (1998) centipede game is a special case of increasing-sum centipede games. Analogous calculations show that the cooperative strategy profile would be the unique optimin point in increasing-sum centipede games with $m'$ or more nodes, where $m'$ depends on the specific payoff function.} This leads to the following result:
	
\begin{claim}
Assume that $k\geq 4$ in the centipede game $(\{1, 2\},$ $\{C,S\}, u_1, u_2, m)$. The strategy profile of always choosing $C$ is the unique optimin.
\end{claim}

The following is well-known: 
\begin{claim}
The strategy profile of always choosing $S$ is the unique SPNE in the centipede game $(\{1, 2\},$ $\{C,S\}, u_1, u_2, m)$.
\end{claim}

Consider also constant-sum centipede games where the sum of the utilities at every outcome equals a constant. By Proposition~\ref{thm:zerosum}, the Nash equilibria coincide with the optimin points because the game is constant-sum. Since player 1 stops at the first node in all Nash equilibria, the optimin criterion uniquely predicts SPNE-type stopping behavior in this class of centipede games.
	
Centipede games have been studied experimentally, starting with \citet{mckelvey1992} and including \citet{fey1996}, \citet{nagel1998}, \citet{rubinstein2007}, and \citet{levitt2011}. One of the most consistently replicated findings is that, on average, subjects show the most cooperative behavior in increasing-sum centipedes and the most non-cooperative (SPNE) behavior in constant-sum centipedes. (For a meta-study of almost all published centipede experiments, see Krockow, Colman, and Pulford, \citeyear{krockow2016}). The direction of these findings is consistent with the optimin criterion: in constant-sum centipedes, the optimin criterion coincides with the unique SPNE prediction, whereas in increasing-sum centipedes, the optimin criterion leads to cooperation when the number of decision nodes is at least four. Moreover, as the number of decision nodes increases, the gap between the performance of cooperative and non-cooperative behavior widens in increasing-sum centipedes, but this gap narrows as the number of decision nodes decreases. Eventually, as the game progresses, the performance of non-cooperative behavior becomes greater than that of cooperation, which is consistent with the decreasing rate of cooperation observed in experiments toward the end of the game.
	
	\subsection{The traveler's dilemma}
	\label{subsec:td}
	
	Figure~\ref{fig:travelers2} illustrates the traveler's dilemma, introduced by \citet{basu1994}. It is a symmetric two-person game in which players pick a number from 2 to 100. The one who picks the lower number receives the dollar amount equal to her choice, plus a \$2 reward, and the other receives a \$2 penalty based on the lower number. If both players choose the same number, they each receive exactly that amount. The payoff function for player $i\in\{1,2\}$ when she plays $a$ and her opponent plays $b$ is defined as follows: for all $a,b$ in $X=\{2,3,...,100\}$,  $u_i(a,b)=\min\{a,b\}+r\times sgn(b-a)$, where $sgn$ denotes the sign function, and $r>1$ determines the magnitude of the reward and penalty. The value of the integer $r$ is set to $2$ in the original game. Regardless of the magnitude of the reward/penalty, the unique Nash equilibrium is $(2,2)$. 
	
	\begin{figure}
		\[
		\begin{array}{ r|c|c|c|c|c| }
			\multicolumn{1}{r}{}
			&  \multicolumn{1}{c}{\text{100}}
			& \multicolumn{1}{c}{\text{99}}
			& \multicolumn{1}{c}{\cdots}
			& \multicolumn{1}{c}{\text{3}}
			& \multicolumn{1}{c}{\text{2}} \\
			\cline{2-6}
			\text{100}&  100,\;100 & 99-r,\;99+r & \cdots & 3-r,\;3+r & 2-r,\;2+r\\
			\cline{2-6}
			\text{99}&  99+r,\;99-r & 99,\;99 & \cdots & 3-r,\;3+r & 2-r,\;2+r\\
			\cline{2-6}
			\text{\vdots}&  \vdots & \vdots & \ddots & \vdots & \vdots \\
			\cline{2-6}
			\text{3}&  3+r,\;3-r & 3+r,\;3-r & \cdots & 3,\;3 & 2-r,\;2+r\\
			\cline{2-6}
			\text{2}&  2+r,\;2-r & 2+r,\;2-r & \cdots & 2+r,\;2-r & 2,\;2 \\
			\cline{2-6}
		\end{array}
		\]
		\caption{Traveler's dilemma with reward/penalty parameter $r$.}
		\label{fig:travelers2}
	\end{figure}
	
	\begin{figure}
		\centering
		\resizebox{0.75\textwidth}{!}{%
        \begin{tikzpicture}
            \begin{axis}[
                width=13cm, height=8cm,
                xlabel={Reward/penalty parameter}, 
                ylabel={Performance},
                x tick label style={/pgf/number format/1000 sep=},
                enlargelimits=0.06,
                xmin=2, xmax=100,
                ymin=-98, ymax=97,
                grid,
                legend pos=north east,
                grid style=dashed
            ]
                \addplot coordinates {
                    (2,97) (7,87) (12,77) (17,67) (22,57) (27,47) (32,37) (37,27)
                    (42,17) (47,7) (52,-3) (57,-13) (62,-23) (67,-33) (72,-43) (77,-53)
                    (82,-63) (87,-73) (92,-83) (97,-93)
                };
                
                \addplot coordinates {
                    (2,87) (7,77) (12,67) (17,57) (22,47) (27,37) (32,27) (37,17)
                    (42,7) (47,-3) (52,-13) (57,-23) (62,-33) (67,-43) (72,-53) (77,-63)
                    (82,-73) (87,-83) (92,-90) (97,-95)
                };
                
                \addplot coordinates {
                    (2,47) (7,37) (12,27) (17,17) (22,7) (27,-3) (32,-13) (37,-23)
                    (42,-33) (47,-43) (52,-50) (57,-55) (62,-60) (67,-65) (72,-70) (77,-75)
                    (82,-80) (87,-85) (92,-90) (97,-95)
                };
                
                \addplot coordinates {
                    (2,7) (7,-3) (12,-10) (17,-15) (22,-20) (27,-25) (32,-30) (37,-35)
                    (42,-40) (47,-45) (52,-50) (57,-55) (62,-60) (67,-65) (72,-70) (77,-75)
                    (82,-80) (87,-85) (92,-90) (97,-95)
                };
                
                \addplot coordinates {
                    (2,2) (7,2) (12,2) (17,2) (22,2) (27,2) (32,2) (37,2)
                    (42,2) (47,2) (52,2) (57,2) (62,2) (67,2) (72,2) (77,2)
                    (82,2) (87,2) (92,2) (97,2)
                };

                \legend{(100,100), (90,90), (50,50), (10,10), (2,2)}
            \end{axis}
        \end{tikzpicture}
		}
		\caption{The effect of reward/penalty parameter on the performance of different strategy profiles for each player in the traveler's dilemma.}
		\floatfoot{Note: The unique symmetric optimin point is to play the highest number when $r$ is small and the lowest number when $r$ is large. The unique Nash equilibrium is to choose the lowest number regardless of $r$.}
		\label{fig:traveler_performance}
	\end{figure}
	
	\begin{claim}
		\label{prop:traveler}
		The performance function for player $i$, if she chooses $a$ and her opponent chooses $b$, is given as follows:
		\begin{equation*}
			\pi_i(a,b)=\begin{cases}
				b-r & \mbox{if } a>b \text{ for } a\in X \\
				a - 2r + 1  & \text{if } a=b,\; a\neq 2, \text{ and } (a +1 - r)\geq 2 \\
				2-r & \text{if } a=b, \;a\neq 2, \text{ and } (a +1 - r)<2\\
				2 & \text{if } a=2 \\
				a  - 3r + 1 & \text{if } a<b, \;a\neq 2, \text{ and } (a +1 - 2r)\geq 2  \\
				2-r & \text{if } a<b, \;a\neq 2, \text{ and } (a +1 - 2r)< 2.
			\end{cases}
		\end{equation*}
	\end{claim}
	
	\begin{proof}
		The first case follows directly from the game's payoff function. The fourth case, $a=2$, is also straightforward. 
		
		Consider the strategy profile $(a,b)$ where $a=b\neq 2$. In this case, each player receives a payoff of $b$. The minimum number player 2 could profitably deviate to is $b - r + 1$, which is bounded from below by 2. As a result of this deviation, player 1's payoff would decrease to $b - 2r + 1$.  If $(a +1 - r)<2$, the minimum number player 2 could profitably deviate to is 2, in which case player 1's payoff would decrease to $2 - r$. 
		
		Now, consider the last two cases with $2<a<b$, where player 1 receives $a+r$ and player 2 receives $a-r$. The minimum number player 2 could profitably deviate to is $a-2r  +1$, which is bounded from below by 2, in which case player 1's payoff would decrease to $a-2r +1 -r$.  If $a-2r +1<2$, the minimum number player 2 could profitably deviate to is 2, in which case player 1's payoff would decrease to $2 - r$, as desired. 
	\end{proof}

    	Observe that for $r<50$, the performance function is Pareto optimal at $(a,b)=(100,100)$ and that for $r\geq 50$ it is Pareto optimal at $(a,b)=(2,2)$. Hence:
	
	\begin{claim}
		The unique symmetric optimin in the traveler's dilemma is $(100,100)$ for $r<50$ and $(2,2)$ for $r\geq 50$.  
	\end{claim}

The following result is well-known: 

	\begin{claim}
	The unique Nash equilibrium in the traveler's dilemma is $(2,2)$ for $r\geq 2$.
\end{claim}
	
	  As noted above, choosing the highest number is the unique symmetric optimin point when $r<50$. As the reward parameter $r$ increases, the performance of the optimin point gradually decreases, as illustrated in Figure~\ref{fig:traveler_performance}. When $r\geq 50$, the unique symmetric optimin point becomes the profile $(2,2)$, which is also the unique Nash equilibrium of the game regardless of the parameter $r$.\footnote{The unique rationalizable strategy profile \citep{bernheim1984,pearce1984} and the correlated equilibrium \citep{aumann1974} coincide with the Nash equilibrium irrespective of $r$.} The optimin criterion is consistent with both the convergence of play to the highest number when the reward is small and the convergence to the lowest number when the reward is large. 
	
	It has been shown in many experiments that players, on average, do not choose the Nash equilibrium strategy, and that changes in the reward/penalty parameter affects observed behavior in experiments. \citet{goeree2001} found that when the reward was high, 80\% of the subjects chose the Nash equilibrium strategy, but when the reward was small, a similar percentage of players chose the highest number. This finding is a confirmation of \citet{capra1999}, in which play converged toward the Nash equilibrium over time when the reward was high, but toward the other extreme when the reward was small. Similarly, \citet{rubinstein2007} found, in a web-based experiment without payments, that 55\% of 2985 subjects chose the highest amount, while only 13\% chose the Nash equilibrium when the reward was small. 
	
	\subsection{The finitely repeated $n$-person public goods game}
	\label{subsec:public_goods}
	
	Another important class of games in economics is public goods games. The main characteristic of these games is that not contributing to the public pot is a dominant strategy for every player. However, if everyone contributes, all players are better off. 
    
    In this subsection, I focus on finitely repeated public goods games, played over $k$ stages, with pure strategies and no discounting. Let the stage-game utility function of player $i$ in an $n$-person (linear voluntary contribution) public goods game be defined as:
	\[
	u_i(a_i,a_{\text{-}i})=\bar{a}-a_i+\frac{m}{n}\sum_{j=1}^{n}a_j,
	\]
	where $\bar{a}\in \mathbb{N}$ denotes the maximum amount each player can contribute, $a_i\geq 0$ is the contribution of player $i$, and $m/n$ is \textit{marginal per capita return} (MPCR). Here, $n\in \mathbb{N}$, $m\in \mathbb{R}$, and the parameters satisfy $n>m>1$. Each stage-game is called a round.
	
	Even in a relatively simple example with $\bar{a}=10, k=10,$ and $n=4$, observe that the number of pure strategy profiles is very large. For this reason, this subsection focuses on comparing two main strategy profiles, whose performances are relatively straightforward to calculate. These profiles provide bounds for the optimin performance, enabling comparative statics with respect to variables such as the number of players and the MPCR.
	
	Define a specific \textit{conditional cooperation strategy} $\bar{\sigma}_i$ as follows: on each round, player $i$ contributes $a_i=\bar{a}$ unless, in some round $k'<k$, another player $j\neq i$ contributes $a_j<\bar{a}$. In that case, player $i$ contributes $a_i=0$ from round $k'+1$ onward (including round $k'+1$). Let $\bar{\sigma}$ denote the conditional cooperation strategy profile. The following claim provides the performance of $\bar{\sigma}$ for each player.
	
	\begin{claim}
		\label{prop:public_goods}
		Consider a $k$-round repeated public goods game with pure strategies. Let $r\leq k$ satisfy $0<r < m(n-1)/n(m-1)$ and $r +1 \geq m(n-1)/n(m-1)$, where $r\in \mathbb{N}$. Then, for each player $i$, the performance of the conditional cooperation profile $\bar{\sigma}$ is given by
		\begin{equation}
			\label{eq:value_cooperation}
			\pi_i(\bar{\sigma})=(k-r)m\bar{a}+\frac{m \bar{a}}{n}+(r-1)\bar{a}.
		\end{equation}
		
	\end{claim}
	
	\begin{proof}
		The proof strategy is to find the earliest round in which a player can unilaterally and profitably deviate from the conditional cooperation profile $\bar{\sigma}$. Once this round is determined, the performance $\pi_i(\bar{\sigma})$ for player $i$ will be calculated based on the worst-case payoff for player $i$, assuming that all other players deviate unilaterally and profitably at this identified round. This approach suffices because deviations in later rounds would allow player $i$ to obtain cooperative payoffs in more rounds.

		To find the round where the first unilateral profitable deviation occurs, fix a player $i$ and focus on round $k - r + 1$, where there are $r \geq 1$ rounds remaining in the game. First, consider the payoff from following the strategy $\bar{\sigma}$ for the remaining $r$ rounds. Since no player has deviated up to round $k - r + 1$, each remaining round will yield a payoff of $m\bar{a}$ to player $i$. Therefore, the total payoff over $r$ rounds is
		\[
		r m \bar{a}.
		\]
		Second, consider $i$'s payoff from deviating by contributing $0$ in round $k - r + 1$. To check if there is a unilateral profitable deviation in this round, all other $n-1$ players continue to contribute $\bar{a}$, so player $i$'s payoff in that round is
		\[
		\frac{m (n - 1)\bar{a}}{n} + \bar{a},
		\]
		where $m(n - 1)\bar{a}/n$ represents player $i$'s payoff from the public good and $\bar{a}$ is the amount saved by not contributing. For the remaining $r - 1$ rounds, every player contributes 0 according to $\bar{\sigma}$, resulting in a payoff of $\bar{a}$ per round for player $i$. Thus, the total payoff from deviating is
		\[
		\left( \frac{m (n - 1)\bar{a}}{n} + \bar{a} \right) + (r - 1)\bar{a}.
		\]
		
		The unilateral deviation to 0 contribution in round $k - r + 1$ is profitable if
		\[
		r m \bar{a} < \frac{m (n - 1)\bar{a}}{n} + \bar{a} + (r - 1)\bar{a}.
		\]
		Simplifying this inequality yields
		\[
		r < \frac{m(n - 1)}{n(m - 1)}.
		\]
		
		Now, in the $k$-round repeated game, let $r\in \mathbb{N}$ satisfy $0<r < m(n-1)/n(m-1)$ and $r +1 \geq m(n-1)/n(m-1)$. This means the earliest profitable deviation for any player occurs at round $k - r + 1$, or equivalently, when there are $r$ rounds remaining.
		
		It is left to calculate the performance $\pi_i(\bar{\sigma})$ for player $i$. The worst-case payoff for player $i$, given unilateral profitable deviations by the other players, occurs when all other players simultaneously deviate to 0 at this earliest round. Specifically, during the first $k - r$ rounds, all players contribute $\bar{a}$, yielding a total payoff of
		\[
		(k - r) m \bar{a}.
		\]
		
		In the deviation round $k - r + 1$, the other $n - 1$ players contribute  $0$, while player $i$  contributes $\bar{a}$. Player $i$'s payoff is therefore
		\[
		\frac{m \bar{a}}{n}.
		\]
		
		In round $k - r + 2$ through $k$, all players contribute $0$, resulting in a payoff of $\bar{a}$ for player $i$ in each of the remaining $r - 1$ rounds. The total payoff for player $i$ from these rounds is
		\[
		(r - 1)\bar{a}.
		\]
		
		Combining these payoffs, the performance for player $i$ under $\bar{\sigma}$ is
		\[
		\pi_i(\bar{\sigma}) = (k - r) m \bar{a} + \frac{m \bar{a}}{n} + (r - 1)\bar{a}.
		\]
	\end{proof}
	
	Equation~\ref{eq:value_cooperation} provides the performance of each player under the conditional cooperation profile $\bar{\sigma}$. To compare, first note the following well-known result:
	
	\begin{claim}
		The unique SPNE in the $k$-round repeated public goods game is the strategy profile where each player contributes 0 in every round regardless of prior contributions. 
	\end{claim}
	
	The performance of the SPNE for player $i$ is simply $i$'s actual payoff, $k\bar{a}$, because no unilateral profitable deviation exists from this \textit{free-riding behavior} of contributing 0 in every round regardless of history.
	
	If the players' performances under the conditional cooperation profile are strictly greater than those under the SPNE, the SPNE cannot be an optimin. In such cases, by Remark~\ref{property:ordinal}, there must exist an optimin whose performance Pareto dominates that of the SPNE. If $\bar{\sigma}$ is not itself an optimin, there must be some optimin that  Pareto dominates the performance of $\bar{\sigma}$. Thus, for each player, Equation~\ref{eq:value_cooperation} provides a lower bound for the performance of such an optimin.
	
	It is straightforward to check that for sufficiently low values of the MPCR, a unilateral profitable deviation from conditional cooperation will occur from the very first round. In this case, the performance of the SPNE cannot be Pareto dominated, and thus the SPNE would be an optimin. The simplest example of this is when $k=1$, where the unique optimin coincides with the unique Nash equilibrium, regardless of the MPCR.

	\begin{figure}
		\centering
		\resizebox{0.65\textwidth}{!}{%
        \begin{tikzpicture}
            \begin{axis}[
                height=7cm,
                width=13cm, 
                xlabel=Number of rounds,
                ylabel=Performance,
                ybar=0.4,
                x tick label style={
                    /pgf/number format/1000 sep=},
                enlargelimits=0.1,  
                ymin=0, ymax=270,
                grid,
                legend pos=north west,
                grid style=dashed,
                xtick={1,2,3,4,5,6,7,8,9,10},  
                xmin=0.5, xmax=10.5             
            ]y
                \addplot[fill=gray!10] coordinates {(1,10)(5,50)(7,70)(10,100)};
                \addplot[pattern=horizontal lines, fill=gray!40] coordinates {(1,3)(5,45)(7,69)(10,105)};
                \addplot[pattern=grid, fill=gray!80] coordinates {(1,8)(5,128)(7,188)(10,278)};

                \legend{SPNE, MPCR = 0.30, MPCR = 0.75}
            \end{axis}
        \end{tikzpicture}		}
		\caption{The performance of the SPNE and conditional cooperation for a player, across different values of MPCR in the repeated public goods game.}
		\floatfoot{Note: The number of players is $n=4$, and the maximum contribution is $\bar{a}=10$.}
		\label{fig:performance_publicgoods}
	\end{figure}
	
	As mentioned, in finitely repeated public goods games, the unique SPNE is to contribute 0 in every round, regardless of parameters such as the MPCR or the number of rounds. However, this theoretical insensitivity sharply contrasts with experimental findings: see, e.g., \cite*{isaac1984}, \citet{andreoni1988}, \citet{ledyard1995}, and \citet{lugovskyy2017}. Experimental research indicates that cooperation (i) is significantly greater in games with a high MPCR compared to those with a low MPCR, (ii) decreases as the end of the game approaches, (iii) restarts when the finitely repeated game is played again, and (iv) is amplified by pre-play communication.
	
	The optimin criterion is consistent with these experimental findings. First, note that the greater the MPCR, the larger the gap between the performance of conditional cooperation and free-riding, as illustrated in Figure~\ref{fig:performance_publicgoods}. For low MPCR values, free-riding behavior satisfies the optimin criterion unless the stage-game is repeated for sufficiently many rounds, allowing cooperation to eventually pay off. For example, if MPCR = 0.27, the performance of conditional cooperation exceeds the performance of SPNE only when $k\geq 20$. In contrast, for high MPCR values, the performance of conditional cooperation exceeds the performance of SPNE even when the game is repeated only a few times.
	
	Second, as the end of the repeated game approaches, the gap between the performance of conditional cooperation and free-riding behavior narrows in the remaining rounds. Eventually, the performance of free-riding behavior becomes greater than that of cooperation. However, if the finitely repeated game is played again, the performance of cooperation at the beginning of the game is generally greater than that of free-riding behavior, which is consistent with the observed `restart' effect. Finally, the finding that pre-play communication increases cooperation is consistent with the tacit agreement interpretation of the optimin criterion. Pre-play communication may help players to agree on cooperation, which generally leads to a higher performance than free-riding behavior.

	\subsection{The finitely repeated PD}
	\label{subsec:pd}
	
	Consider the finitely repeated PD played over $k>0$ stages, with pure strategies and no discounting. Assume that $T>R>P>S$, where $T$ stands for temptation payoff, $R$ the reward, $P$ the punishment, and $S$ the sucker's payoff (normalized to 0). The unique optimin point in the one-shot game is (Defect, Defect). However, new optimin points emerge when the game is repeated. The stage-game's payoff matrix is presented below.
	
	\[
	\begin{array}{ r|c|c| }
		\multicolumn{1}{r}{}
		&  \multicolumn{1}{c}{\text{Cooperate}}
		& \multicolumn{1}{c}{ \text{$\;$ Defect $\;$} }\\
		\cline{2-3}
		\text{Cooperate}&  R,R & S,T \\
		\cline{2-3}
		\text{Defect}&  T,S & P,P \\
		\cline{2-3}
	\end{array}
	\]
	
	The following is a well-known result:
		\begin{claim}
		The unique SPNE in a $k$-round repeated PD is the strategy profile where players choose DD every round regardless of prior choices. 
	\end{claim}

The next claim follows from the fact that there is no unilateral profitable deviation from the SPNE.
\begin{claim}
	In a $k$-round repeated PD, the performance of the SPNE for each player is $kP$.
\end{claim}

	In the Tit-for-Tat (TFT) strategy, a player starts by cooperating in the first round and, in every subsequent round, mimics the opponent's action in the previous round. TFT is one of the most well-known forms of conditional cooperation in the PD. Another conditional cooperation strategy is the grim trigger strategy, which is stricter than TFT. In the grim trigger strategy, a player cooperates until a defection is observed and then defects for the remainder of the game. The following two claims establish that, under specific conditions, the TFT profile is an optimin.

	\begin{claim}
		Consider a $k$-round repeated PD with pure strategies. Assume that the following conditions are satisfied:
		\[
		\text{ (i) }~ 2R\geq T+P ~\text{and (ii) }~ 3R \geq 2T.
		\] 
		Then, for each player $i$, the performance of the TFT strategy profile performance is $(k-1)R$.
	\end{claim}
	
	\begin{proof}
		The proof strategy is to show that if both (i) and (ii) are satisfied, there can only be a `one-round' unilateral profitable deviation for each player from the TFT profile. As a result of this deviation, the non-deviator's payoff decreases to $(k-1)R$.
		
		Assume that conditions (i) and (ii) are satisfied, and consider the following three cases of unilateral deviations from the TFT profile: (1) deviations involving two rounds, (2) deviations involving three rounds, and (3) deviations involving four or more rounds. 
		
		Case (1): In two rounds, the only potentially profitable unilateral deviation from the TFT profile is to deviate to D in one round and continue to play D in the next round. In these two rounds, the stage-game action profiles are DC and DD, so the deviator's payoff from these rounds would be $T + P$. Condition (i), $2R \geq T + P$, ensures that such a two-round unilateral deviation is not profitable.
		
		Case (2): In three rounds, the only potentially profitable unilateral deviation from the TFT profile is to deviate to D in one round, play C in the next round, and then play D again in the following round. In these three rounds, the stage-game action profiles are DC, CD, and DC. The deviator's payoff from these rounds would be $T + 0 + T = 2T$. Condition (ii), $3R \geq 2T$, ensures that such a three-round unilateral deviation is not profitable.
		
		Case (3): Now consider any unilateral deviation involving $\hat{k} \geq 4$ rounds from the TFT profile. Let $1< k' \leq \hat{k}$ be the number of rounds in which the deviator receives $T$. Since TFT is retaliatory, for every payoff $T$ in a round, the deviator must receive 0 in another round unless $\hat{k}$ is odd. In that case, the deviator may receive $T$ in one additional round, resulting in $ k'-1$ rounds with $0$. This implies that player $i$'s payoff following this deviation would be $k'T + (k'-1)0=k'T$, whereas $i$'s payoff from sticking with the TFT would be at least $k'R+(k'-1)R$. By condition (ii),  $(2k' - 1)R \geq k'T$ for $k'>1$. Thus, player $i$'s deviation does not yield a higher payoff for these $2k' - 1$ rounds out of $\hat{k}$ rounds. In the remaining rounds, the deviator may receive either $R$ or $P$. Since $R > P$, player $i$ has no unilateral profitable deviation involving $\hat{k} \geq 4$ rounds.
		
		Since conditions (i) and (ii) ensure that no unilateral deviations from the TFT profile involving two or more rounds are profitable, each player $i$'s performance from the TFT profile is $(k - 1)R$, which results from a single-round profitable deviation by the opponent.
	\end{proof}
	
	\begin{claim}
		Assume that, in addition to the conditions (i) and (ii), the following condition is satisfied: (iii) $k \geq R/(R-P)$. Then, the TFT strategy profile is an optimin in the $k$-round repeated PD.
	\end{claim}
	
	\begin{proof}

The proof strategy is to rule out the relevant strategy profiles that could potentially Pareto dominate the TFT profile's performance. Under the conditions (i) and (ii), the only profiles that need to be checked are those that yield each player either strictly more than $(k - 1)R$ or exactly $(k - 1)R$ for one player and more for the other. These profiles are classified based on their path of play action profiles (i.e., CC, DD, CD, or DC) that could generate sufficiently high payoffs. In each case, it is shown that either there is a profitable deviation for one player that decreases the other's payoff strictly below $(k - 1)R$, or each player's performance is exactly $(k - 1)R$. This eliminates any possibility of Pareto domination over the TFT profile, which shows that it is an optimin.

First, note that condition (iii), which simplifies to $(k - 1)R \geq kP$, rules out the SPNE, as it is impossible for both players to receive payoffs higher than their TFT performances.
		
Next, let $\sigma$ be a pure strategy profile. There are two types of strategy profiles whose performance may potentially Pareto dominate the TFT profile's performance. Either $u_i(\sigma) > (k - 1)R$ for each $i$, or there exists some $i\neq j$ such that $u_i(\sigma) = (k - 1)R$ and $u_j(\sigma) > (k - 1)R$. The following three cases cover the first possibility, while the fourth case covers the second possibility. 
		
Case (1): The path of $\sigma$ contains only the action profiles DD and CC. Let $k''$ be the number of rounds in which DD is chosen. Then, for each $i$, $u_i(\sigma) = (k - k'')R + k''P$.\footnote{It is possible that $u_i(\sigma)>(k - 1)R$ if, for example, $2P > R$ and $k'' = 2$.}  One of the players has a unilateral profitable deviation from the profile $\sigma$ in one of the CC rounds, in which case the non-deviator's payoff would decrease to 0 in that round. Thus, player $i$'s performance from this profile falls strictly below $(k - 1)R$.
		
		Case (2): The path of $\sigma$ contains only CCs except in two rounds, in which players choose CD and DC. In that case, for each $i$, $u_i(\sigma) = (k - 2)R + 0 + T$. Player $j \neq i$, who plays C in the last round, has a unilateral profitable deviation to D, in which case player $i$'s payoff would decrease to $P$. Since $(k - 2)R + 0 + P<(k - 1)R$, player $i$'s performance from this profile falls again strictly below $(k - 1)R$. 
		
		Case (3): Players always choose CC on the path of $\sigma$. In that case, for each player $i$, $u_i(\sigma) = kR$. Then, clearly, there is a unilateral profitable deviation in the last round as in the TFT case, so a player's performance from $\sigma$ cannot be strictly greater than $(k - 1)R$.
		
		To see why the above cases are exhaustive for the first possibility in which both players' payoffs exceed $(k - 1)R$, note that if there is a DC on $\sigma$'s path, for player 2 to exceed a payoff of $(k - 1)R$, there must also be a CD on its path. This is because even if CC is played in all but one rounds, player 2 would receive at most a payoff of $(k - 1)R$. Similarly, if there is a CD on $\sigma$'s path, there must also be a DC. However, if there are both CD and DC on its path, there cannot be a DD, and players each exceed the payoff of $(k - 1)R$. This follows because in those three rounds (CD, DC, and DD), a player receives $T + P + 0$; however, $2R \geq T + P$ by condition (i).
		
		Case (4): For some player (say, player 1), $u_1(\sigma) = (k - 1)R$ and $u_2(\sigma) > (k - 1)R$. There are two subcases: (a) either $\sigma_2$ is not a best-response to $\sigma_1$, or (b) it is a best-response to $\sigma_1$.  In subcase (a), player $2$ has a unilateral deviation that reduces player $1$'s payoff to strictly below $(k - 1)R$. In subcase (b), I show that player 1 has a profitable deviation in the last two rounds that reduces player $2$'s payoff to strictly below $(k - 1)R$. 

        Let $r, p, t_1$, and $t_2$ be the number of CC, DD, DC, and CD action profiles, respectively, on $\sigma$'s path. Case (4) implies that, for $r+ p+t_1+t_2=k$, 
        \[
			 pP+t_1 T + rR = (k-1)R,
		\]
         \[
			 pP+t_2 T + rR >(k-1)R.
	\]
        It follows that $pP+t_2 T + rR > pP+t_1 T + rR$, which simplifies to $t_2>t_1$. This implies that there must be at least one CD. Since player 2 is playing a best response, the last action profile on $\sigma$'s path must be CD. It cannot be DD because, if $\sigma$ satisfies the payoff condition in Case (4), player 2 would have a profitable deviation in the penultimate round.

        Now, consider the first $k-1$ rounds. Let $r', p', t'_1$, and $t'_2$ be the number of CC, DD, DC, and CD action profiles, respectively, on $\sigma$'s path during these $k-1$ rounds. Since player 1 and player 2 receive 0 and $T$, respectively, from the last round, it follows that  
        \[
			 p'P+t'_1 T + r'R = (k-1)R,
	\]
       \[
			 p'P+t'_2 T + r'R >(k-1)R-T.
	\]
        This simplifies to $t'_2+1>t'_1$, which implies that there are at least as many CDs as DCs within these $k-1$ rounds. Player 1 receives only $T$ from any two rounds that include a DC and a CD. However, in the first $k-1$ rounds, player 1 must receive at least $R$ per round on average. This is impossible if there is a DC (i.e., $t'_1>0$) on the path, as $2R>T$ by condition (i). Thus, $\sigma$'s path cannot include a DC. This further implies that it cannot include a CD either. If it did, player 1 would receive 0 in at least one round. To compensate for this loss, there would have to be a corresponding DC, which is impossible as shown above.

        Therefore, $\sigma$'s path must consist of CCs and a CD in the last round. This implies that player 1 can profitably deviate from $\sigma$ by playing D in both the penultimate and last rounds. As a result, player 2's payoff decreases to $(k-2)R+P$, which is strictly below the TFT profile's performance, $(k-1)R$. 
		
	As a result, there is no strategy profile whose performance Pareto dominates the TFT profile's performance. As desired, the TFT profile is an optimin.
	\end{proof}
	
	\begin{cor}
		If conditions (i) -- (iii) are satisfied, then the grim-trigger strategy profile is an optimin.
	\end{cor}
	
	Figure~\ref{fig:performance_TFT} illustrates the performance of the TFT profile and the SPNE in a $k$-round repeated PD, without assuming conditions (i)--(iii). The stage-game payoff matrix is provided below. 
	\[
	\begin{array}{ r|c|c| }
		\multicolumn{1}{r}{}
		&  \multicolumn{1}{c}{\text{Cooperate}}
		& \multicolumn{1}{c}{ \text{$\;$ Defect $\;$} }\\
		\cline{2-3}
		\text{Cooperate}&  3,3 & 0,T \\
		\cline{2-3}
		\text{Defect}&  T,0 & 1,1 \\
		\cline{2-3}
	\end{array}
	\]
	\noindent The performance of SPNE does not depend on $T$, whereas the performance of the TFT profile decreases as $T$ increases. When $T$ becomes very large (e.g., $T=22$), the performance of the SPNE exceeds the performance of the TFT profile. However, this does not necessarily imply that the SPNE is an optimin, as there may be another strategy profile whose performance Pareto dominates that of the SPNE.\footnote{When $2R<T $, a different cooperative strategy profile may emerge whose performance Pareto dominates the TFT profile's performance. For example, assume that the payoffs are $T=10$, $R=3$, and $P=1$. Consider a strategy profile where players alternate between DC, CD, DC, CD, $...$, unless a deviation occurs, in which case they play $D$ indefinitely thereafter. Notice that this profile would have a greater performance than the usual TFT profile.}
	
	\begin{figure}
		\centering
		\resizebox{0.7\textwidth}{!}{%
		\begin{tikzpicture}
			\begin{axis}[
				width=12cm, height=8cm,
				xlabel={Number of rounds}, 
				ylabel={Performance},
				xticklabel style={/pgf/number format/1000 sep=},
				enlargelimits=0.06,
				xmin=0, xmax=10,
				ymin=0, ymax=28,
				grid,
				legend pos=north west,
				grid style=dashed,
			]
				\addplot coordinates {
					(1,1) (2,2) (3,3) (4,4) (5,5) (6,6) (7,7) (8,8) (9,9) (10,10)
				};
				\addplot coordinates {
					(1,0) (2,3) (3,6) (4,9) (5,12) (6,15) (7,18) (8,21) (9,24) (10,27)
				};
				\addplot coordinates {
					(1,0) (2,1) (3,2) (4,5) (5,8) (6,11) (7,14) (8,17) (9,20) (10,23)
				};
				\addplot coordinates {
					(1,0) (2,1) (3,2) (4,3) (5,4) (6,5) (7,6) (8,7) (9,8) (10,9)
				};
				\legend{SPNE,$T=4$,$T=9$,$T=22$}
			\end{axis}
		\end{tikzpicture}
		}
		\caption{The performance of the SPNE and the TFT profile for a player, across different values of $T$, in the repeated PD.}
		\floatfoot{Note: $R=3, P=1$, and $S=0$.}
		\label{fig:performance_TFT}
	\end{figure}

	The literature on finitely repeated PD games is extensive: see, for example, \citet{axelrod1980}, \citet{selten1986}, recent meta-studies by \citet{mengel2017}, \citet*{embrey2017}, and the references therein. It is well established that players cooperate more frequently than predicted by the SPNE. More specifically, (i) initial cooperation becomes more likely as the number of rounds increases, and (ii) cooperation declines as the end of the game approaches. The optimin criterion is consistent with these regularities. Cooperation generally satisfies the optimin criterion in the repeated PD because, even if a player tries to exploit cooperative behavior, the cooperator's minimal payoff is greater than the SPNE payoff. As the number of rounds in the game increases, the performance of cooperation also increases. However, the performance of cooperation gradually decreases as the game progresses, eventually falling below that of defection. This is consistent with the finding that cooperation levels decrease toward the end of the game.
	
	\section{Relevant literature}
	\label{sec:literature_review}
	
	The closest concept to the optimin criterion is that of the maximin criterion, which has been studied in different contexts by a number of researchers, including \citet{borel1921}, \citet{neumann1928}, \citet{wald1939}, and \citet{rawls1971}. There are also axiomatizations of the maximin criterion, including by \citet{milnor1954} and \citet{gilboa1989}. For a related axiomatization, see \citet{puppe2009}, who show that the axioms of \citet{milnor1954} that characterize the maximin decision rule are consistent with ignoring some states in which all payoffs are `small.' In their seminal work, \citet{gilboa1989} characterize maxmin expected utility using a set of intuitive axioms for cautious decision-makers with a set $C$ of (multi-prior) subjective beliefs. In the Gilboa-Schmeidler maxmin model, the set $C$ can be varied to rationalize different choices; similarly, in the optimin model, a constraint $B_{\text{-}i}$ can be varied to rationalize different choices in a game setting. Since this paper explicitly considers mixed strategies, the comparison is especially pertinent to pure optimin when the framework is restricted to pure strategies. For a related discussion, see \citet{kuzmics2017}.
	
	The literature on solution concepts that incorporates various levels of cautiousness in games includes \citet{selten1975}, \citet{basu1991}, and, more recently, \citet{perea2006}, \citet{renou2010}, and \citet*{iskakov2018}. Prominent equilibrium concepts under ambiguity include \citet{dow1994}, \citet{lo1996}, \citet{klibanoff1996}, \citet{marinacci2000}, and more recently \citet{azrieli2011}, \citet{bade2011}, \citet{riedel2014}, and \citet{battigalli2015}. For an overview of the ambiguity models in games, see \citet{mukerji2004} and \citet{beauchene2014}.
	
	This paper also contributes to the literature on theoretical explanations for experimental deviations from Nash equilibrium toward cooperation. Prominent models in the literature typically explain systematic deviations by focusing on social preferences \citep[e.g.,][]{klumpp2012}, cognitive hierarchy of players \citep*[e.g.,][]{stahl1993,nagel1995,camerer2004}, dynamic reasoning \citep[e.g.,][]{brams1994}, reciprocity \citep[e.g.,][]{ambrus2011}, common knowledge \citep[e.g.,][]{aumann1992,binmore1994}, incomplete information, bounded rationality \citep[e.g.,][]{radner1980,radner1986,mckelvey1995}, preferences over strategies \citep[e.g.,][]{segal2007}, repeated games with random matching \citep[e.g.,][]{kandori1992,heller2017}, and learning \citep[e.g.,][]{mengel2014}. Some of the early works include \citet{kreps1982a}, \citet{kreps1982b}, \citet{sobel1985}, \citet{Fudenberg1986}, and \citet{neyman1999}, who show that incomplete information about various aspects of a game can explain cooperation in finitely repeated games. \citet{neyman1999} shows that cooperation can be induced in finitely repeated PD if players are restricted to using strategies with bounded complexity. Non-utility-maximizing models include imitation-based decision-making in games \citep*[e.g.,][]{eshel1998}. For relevant literature on experiments and more details about theoretical explanations, see section~\ref{sec:apps} and the references therein.
	
	The optimin criterion differs from the aforementioned models and solution concepts along three main dimensions: (i) conceptual/cognitive background, (ii) scope of application, and (iii) predictive power. First, the optimin criterion is a non-equilibrium concept wherein players maximize their minimal payoffs subject to the constraint of unilateral profitable deviations. This criterion offers a novel extension of maximin reasoning from two-person zero-sum games to $n$-person non-zero-sum games. Second, while the optimin criterion mainly applies to non-cooperative games, it can also be defined in cooperative games and matching models, which I explore in a companion paper. Third, in contrast to other solution concepts, the optimin criterion is consistent with the direction of non-Nash deviations in the games analyzed in this paper.
	
	In conclusion, it seems unlikely that complex human behavior can be fully captured by a single solution concept. For example, the (11-21)-type games introduced by \citet{arad2012} seem to naturally invoke level-$k$ reasoning \citep{stahl1993}, whereas in complex games, such as Blotto games, players---who are unable to calculate optimal strategies---focus on the characteristics of strategies rather than the strategies themselves (for a formalization of this type of reasoning, see Arad and Rubinstein, \citeyear{arad2019}). However, a solution concept and a `reasoning process' are generally two distinct notions. For instance, in the context of Nash equilibrium, there has been a longstanding question about whether any reasonable process can lead players to arrive at a Nash equilibrium. Evolutionary game theory has provided some positive answers to this question in certain classes of games (see, e.g., Weibull, \citeyear{weibull1995}, and Hofbauer and Sigmund, \citeyear{hofbauer1998}); however, there are also impossibility results (see, e.g., Chapter 8 of Hofbauer and Sigmund, \citeyear{hofbauer1998}, and Hart and Mas-Colell, \citeyear{hart2003}, \citeyear{hart2006}). Regarding optimin points, a promising direction for future research is to explore the reasoning processes that might converge to an optimin.

	\section*{Appendix}
	\label{appendix}
	
	\subsection*{Berge's theorem}
		
	I begin by stating Theorem 1 from \citet[p. 115]{berge1959}, as presented by \citet[p. 569]{aliprantis1994}, which is key to the proof of the existence result. This theorem is one of the two main results used in the proof of Berge's famous Maximum Theorem. Throughout, I adopt the terminology used by Aliprantis and Border.
	
	\noindent\textbf{Lemma 17.29} (\citeauthor{aliprantis1994}).
	\textit{Let  $\varphi: X \rightrightarrows Y$  be a lower hemi-continuous correspondence between topological spaces, and let the function  $f : Gr(\varphi) \to \mathbb{R}$ be lower semi-continuous. Define the extended real function  $m: X \to \mathbb{R}^*$ by
		\[
		m(x) = \sup_{y \in \varphi(x)} f(x, y),
		\]
		where as usual,  $\sup \emptyset = -\infty$. Then the function $m$ is lower semi-continuous.}

	\subsection*{Proof of Lemma 1}
	The following lemma presents a property of the performance function which is used in Theorem~\ref{thm:existence}.
	
	\begin{lem}[Upper semi-continuity]
		\label{value_usc}
		For every player $i \in N$, the performance function $\pi_i: S\rightarrow \mathbb{R}$ is upper semi-continuous.
	\end{lem}
	
	\begin{proof}
		In six main steps, I show that the performance function $\pi_i$ of player $i$ in a two-player game $(S_i, u_i)_{i\in \{1,2\}}$ is upper semi-continuous. Steps 1 to 3 involve decomposing the performance function of player $i$, establishing the lower hemi-continuity of the strict better-response correspondence of player $j$, and ensuring the appropriate domain conditions are met to apply Berge's theorem. Step 4 directly applies Berge's theorem to the setting. Finally, Steps 5 and 6 show that the decomposed performance function is upper semi-continuous by showing it is the minimum of two upper semi-continuous functions.
		
		First, let $\tilde{B}_j$ denote the strict better-response correspondence of player $j\in \{1,2\}$, defined as follows: for every $p\in S$,
		$\tilde{B}_j(p)=\{p'_j\in S_j\mid u_j(p'_j,p_{\text{-}j})>u_j(p_j,p_{\text{-}j})\}$. Since for each player $j\in \{1,2\}$, $B_j(p)=\tilde{B}_j(p)\cup \{p_j\}$, the performance function can be decomposed as follows. 
		For every player $i\in \{1,2\}$, with $i\neq j$, and all $p \in S$, 
		\begin{equation}
			\label{eq:decomposition}
			\pi_i(p)=\min\{\inf_{p'_{j}\in \tilde{B}_j(p)} u_{i}(p_i,p'_{j}),\,u_{i}(p)\}.
		\end{equation}
		To see this, note that if, for a given $p$, the infimum in the performance function (see Definition~\ref{def:value}) is attained at some $p'_{j}\in \tilde{B}_j(p)$, then Equation~\ref{eq:decomposition} holds.\footnote{If the strict better-response correspondence is empty for some $p$, then $\pi_i(p) = u_i(p)$. The right-hand side of  Equation~\ref{eq:decomposition} also simplifies to $u_i(p)$ since the infimum over an empty set is plus infinity.} If it is not attained at any $p'_{j}\in \tilde{B}_j(p)$, then $j$ has a unilateral profitable deviation that decreases $i$'s payoff: $\inf_{p'_{j}\in \tilde{B}_j(p)} u_{i}(p_i,p'_{j})<u_{i}(p)$. This implies that 
		\[
		\inf_{p'_{j}\in \tilde{B}_j(p)\cup \{p_j\}} u_{i}(p_i,p'_{j}) = \inf_{p'_{j}\in \tilde{B}_j(p)} u_{i}(p_i,p'_{j}).
		\]
		Hence Equation~\ref{eq:decomposition} follows.
		
		Second, I show that for every $j\in \{1,2\}$, where $j\neq i$, the strict better-response correspondence $\tilde{B}_j:S_i\times S_j \rightrightarrows S_j$ is lower hemi-continuous. To prove this, it is sufficient to show that the graph of $\tilde{B}_j$ defined as follows is open.
		\[
		Gr(\tilde{B}_j)=\{(p_j,q)\in S_j\times S_i\times S_j\mid p_j\in \tilde{B}_j(q)\}.
		\]
		$Gr(\tilde{B}_j)$ is open in $S_j\times S_i\times S_j$ if and only if its complement is closed. Let $(p^k_j,q^k_i,q^k_j)^{\infty}_{k=1}$ be a sequence in the complement of the graph of $\tilde{B}_j$, 
		\[[Gr(\tilde{B}_j)]^\mathsf{c}=(S_j\times S_i\times S_j) \setminus Gr(\tilde{B}_j),
		\]
		converging to $(p_j,q_i,q_j)$, where  for all $k$, $p^k_j\notin \tilde{B}_j(q^{k})$. In other words, for all $k$, $u_{j}(p^k_j,q^k_{i})\leq u_{j}(q^k)$. Since $u_i$ 
		bilinear, it is continuous, which implies that $u_{j}(p_j,q_{i})\leq u_{j}(q)$, meaning $p_j\notin \tilde{B}_j(q)$. Hence $[Gr(\tilde{B}_j)]^\mathsf{c}$ is closed, implying that  $\tilde{B}_j$ is lower hemi-continuous.
		
		Third, define $\hat{u}_i:S_j\times S_i\times S_j \rightarrow \mathbb{R}$ as follows: for all $(p_j,q_i,q_j)\in S_j\times S_i\times S_j$, $\hat{u}_i(p_j,q_i,q_j)=u_i(p_j,q_i)$. Since $u_i$ is continuous, it follows that $\hat{u}_i$ is also continuous. In addition, define $\bar{u}_i:Gr(\tilde{B}_j) \rightarrow \mathbb{R}$ as the restriction of $\hat{u}_i$ to $Gr(\tilde{B}_j)$, that is $\bar{u}_i=\hat{u}_{i_{|Gr(\tilde{B}_j)}}$. The continuity of $\hat{u}_i$ implies the continuity of its restriction $\bar{u}_i$, which, in turn, implies that the negative of the function, $-\bar{u}_i:Gr(\tilde{B}_j)\rightarrow \mathbb{R}$, is lower semi-continuous.
		
		Fourth, to apply Theorem 1 of \citet{berge1959} (see Lemma 17.29 from Aliprantis and Border, \citeyear{aliprantis1994}), define the function $m_i:S_i\times S_j \rightarrow \mathbb{R}$ by 
		\begin{equation}
			\label{eq:berge}
			m_i(q)=\sup_{p_j\in \tilde{B}_j(q)}-\bar{u}_i(p_j,q).
		\end{equation}
		Since the strict better-response correspondence $\tilde{B}_j$ is lower hemi-continuous and the function $-\bar{u}_i:Gr(\tilde{B}_j)\rightarrow \mathbb{R}$ is lower semi-continuous, by Berge's Theorem 1, the function $m_i$ is lower semi-continuous.
		
		Fifth, multiplying both sides of Equation~\ref{eq:berge} by $-1$ and using the duality of supremum and infimum imply that for every $q\in S$, Equation~\ref{eq:sup} holds:
		\begin{align}
			-m_i(q) &=  \inf_{p_j\in \tilde{B}_j(q)}\bar{u}_i(p_j,q) \label{eq:sup} \\
			&= \inf_{p_j\in \tilde{B}_j(q)}u_i(p_j,q_i). \label{eq:sup2}
		\end{align}
		Equation~\ref{eq:sup2} follows from the definitions of functions $\hat{u}_i$ and $\bar{u}_i$. Now, define the function $\tilde{\pi}_i:S_i\times S_j \rightarrow \mathbb{R}$ such that for every $q\in S$, $\tilde{\pi}_i(q)=\inf_{p_j\in \tilde{B}_j(q)}u_i(p_j,q_i)$. Since the function $m_i$ is lower semi-continuous, the function $\tilde{\pi}_i=-m_i$, is upper semi-continuous.
		
		Finally, note that the performance function in Equation~\ref{eq:decomposition} can be written as follows: for all $q\in S$, $\pi_i(q)=\min\{\tilde{\pi}_i(q),u_{i}(q)\}$. This implies that the performance function $\pi_i$ of player $i$ is upper semi-continuous because the minimum of two upper semi-continuous functions is also upper semi-continuous.
		
		The extension of the proof to the $n$-person case follows analogously  because the key steps, including the lower hemi-continuity of the better-response correspondence, the continuity of utility functions, and the Berge's theorem, hold for any finite number of players.
	\end{proof}

\textit{DATA AVAILABILITY STATEMENT} The code and data that support the illustrations in this study are openly available in Optimin Solver at \url{https://github.com/drmehmetismail/Optimin-Solver}.

\end{document}